\documentclass{article}

\usepackage[cp1250]{inputenc}
\usepackage{graphicx}
\usepackage{float}

\usepackage{mathtools}   
\usepackage{amsfonts}
\usepackage{url}
\usepackage{hyperref}

\newtheorem{Theorem}{Theorem}

\begin{document}

{\LARGE\centering{\bf{Simple explanation of Landauer's bound and its ineffectiveness for multivalued logic}}}

\begin{center}
\sf{Rados\l aw A. Kycia$^{1,3,a}$}\\
\sf{Agnieszka Niemczynowcz}$^{2,b}$
\end{center}

\medskip
\small{
\centerline{$^{1}$Masaryk Univeristy}
\centerline{Department of Mathematics and Statistics}
\centerline{Kotl\'{a}\v{r}sk\'{a} 267/2, 611 37 Brno, The Czech Republic}
\centerline{\\}
\centerline{$^{2}$University of Warmia and Mazury}
\centerline{Faculty of Mathematics and Computer Science}
\centerline{S\l{}oneczna 54, 10-710 Olsztyn, Poland}
\centerline{\\}
\centerline{$^{3}$Cracow University of Technology}
\centerline{Faculty of Materials Science and Physics}
\centerline{Warszawska 24, Krak\'ow, 31-155, Poland}
\centerline{\\}

\centerline{$^{a}${\tt
kycia.radoslaw@gmail.com}, $^{b}${\tt niemaga@matman.uwm.edu.pl}}
}

\begin{abstract}
\noindent
We discuss, using recent results on the Landauer's bound in multivalued logic, the difficulties and pitfalls of how to apply this principle. The presentation is based on Szilard's version of Maxwell's demon experiment and use of equilibrium Thermodynamics. Different versions of thermodynamical/mechanical memory are presented - one-hot encoding version and the implementation based on reversed Szilard's experiment. Relation of the Landauer's principle to Galois connection is explained in detail.
\end{abstract}
Keywords: Landauer's principle, Entropy, Multivalued logic, Encoding, the Second Law of Thermodynamics, thermodynamic memory implementation, Galois connection;  \\

\section{Introduction}
In thermodynamics Maxwell's demon paradox was a long unresolved problem until the 60s when R. Landauer postulated \cite{Landauer} the bound for heat $Q$ emitted during the erasure of one bit of information to be no less than the Landauer's bound $k_{B}T\ln(2)$, where $k_{B}$ is the Boltzmann constant and $T$ is the temperature of the environment in which the memory is embedded. Then it was the suggestion of Bennet \cite{Bennet, BennetDemon} that this bound can be applied to the demon's memory in the cycle to make the Second Law of Thermodynamics applicable to Maxwell's demon experiment and to resolve this long-lasting paradox. 

Currently, the Landauer's principle is under substantial experimental check \cite{LandauerMeasurment}, including quantum level \cite{QuantumLandauer}, as well as, understood on the level of classical (equilibrium) Thermodynamics, where it is equivalent to the Second Law of Thermodynamics \cite{OverviewLandauer, Bennet, BennetDemon}, on the ground of statistical physics \cite{Piechocinska, LandauerExplainedFull} or under theoretical generalizations \cite{LandauerExplained, LandauerExplainedFull, ThermodynamicCostOfDataProcessing} and even abstract formulations using category theory \cite{KyciaLandauer, KyciaLandauer2} with much potential application. Therefore the Landauer's principle is solid stated law.

As it was pointed out in many places (for an excellent review see \cite{ThirdBase}) the ternary system can be seen as a more optimal coding base for numbers than the classical binary system used in contemporary computers. It results from the maximization of the expression $\frac{ln(B)}{B}$, which is average information per digit in the system \cite{InformationEntropy} and associates information for a number system of $B$ letters (digits) per one element of the alphabet. The extremal value is reached at $B=e$ - the base of the natural logarithm. Since $B=3$ is closer to $e$ than $B=2$; therefore, it is suggested that the trit-base system is closer to optimal encoding.

However, in nature, we are also accustomed to the systems with larger than three base, e.g., in living organisms or DNA computing, the 'bits' of DNA or RNA consisting of four fundamental chemical components. In addition, in human culture the systems based on $B=10$, $B=12$, $B=16$ or even $B=60$ are common.

There was a recent attempt to merge the Landauer's principle with non-binary base memory systems. In \cite{TritLandauer}, it was presented that Landauer's principle is applicable for trit memory. The physical bound for trit is $k_{B}T\ln(3)$, however, the correcting factor from the efficiency of coding $B=3$ system in the binary system is $log_{2}(3)$, which restores the original Landauer's bound. It was also suggested that the Szilard version \cite{Szilard} of Maxwell's demon may be used as a memory model also for trit, which deserves much more elaboration and extension to multivalued logic. 

In this paper, we want to explicitly explain with full details, using the Szilard's approach to Maxwell's demon experiment, that the Landauer's bound is valid for a system with an arbitrary base. The presentation is provided using the quasistatic setup of classical thermodynamics for simplicity. We want to strongly underline that all the classical results on Szilrad's realization of Maxwell's demon were throughout explored in various setups, and we do not claim we present something new. We present new insight on the use of the different base of logic in memory and its connection with Landauer's principle, an issue which was recently risen in \cite{TritLandauer}. We also present an insight on the efficiency of coding and its connection with Landauer's bound, as well as some of the practical examples of realization of Galois connection in memory systems that were also recently proposed \cite{KyciaLandauer, KyciaLandauer2}. We think that this adds a better understanding of the Landauer's principle in multivalued logic. 

The paper is organized as follows: First, we present the original Szilard's experiment and apply it for bits. Then we show, by the Second Law of Thermodynamics, how an extension of this experiment can be applied to deduce Landauer's bound. We will also discuss some potential problems in the naive application of Landauer's principle to multivalued logic. Finally, it is presented how mechanical memory can be implemented and at which stage of this implementation, the heat is generated for irreversible operations.

\section{Setup for bit}

\begin{figure}
\centering
 \includegraphics[width=0.5\textwidth]{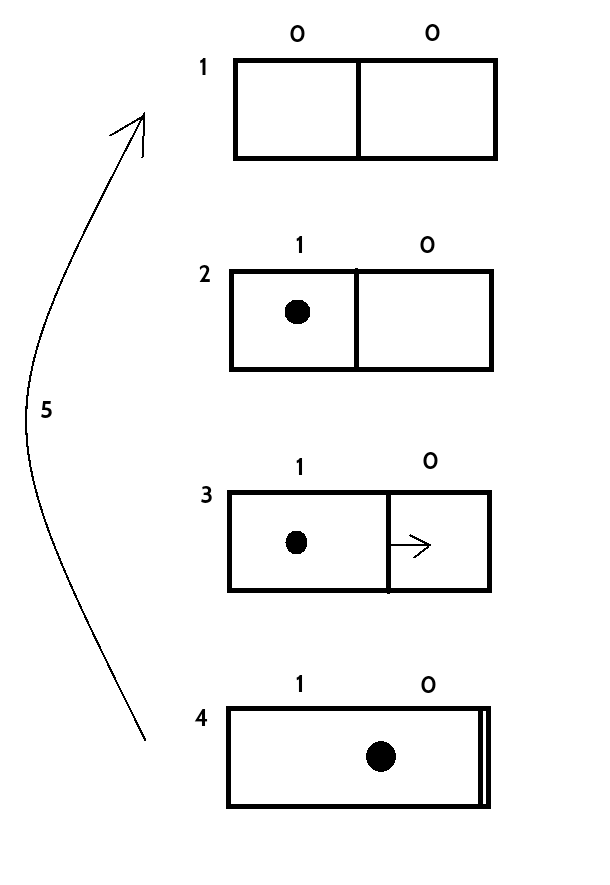}
 \caption{The Szilard's version of Maxwell's demon.}
 \label{Fig.SzilardVersion}
\end{figure}

The standard Szilard's version of Maxwell's demon for binary computations \cite{Szilard, KyciaLandauer, TritLandauer} consists of a box with a single particle of an ideal gas that fulfills the equation of state $pV=k_{B}T$, which is a one-particle version of the equation $pV=nRT$, where $n$ is the number of moles of gas and $nR=Nk_{B}$, where $N$ is the number of particles. It is in the thermal bath of the box at a temperature $T$.  The 'demon' is the additional device that schedules the cycle. The situation is presented in Fig. \ref{Fig.SzilardVersion}. 
The cycle consists of the following steps:
\begin{enumerate}
 \item {The border is put in the middle of the box, splitting the whole volume $V$ into halves. The state of the memory of 'the demon' is erased in the state $[0,0]$.}
 \item {The particle is localized in, e.g., the left part of the box. Here 'the demon', registers the state of the box in its memory by changing its state to $[1,0]$.}
 \item {In this step the border becomes movable, and the gas expands from the volume $V/2$ to $V$ isothermally giving the work $W=\int_{V/2}^{V}k_{B}T \frac{dV}{V}=k_{B}Tln(2)$.}
 \item {When the border reach the right end of the box the extraction of the work stops.}
 \item {In order to return to the initial step, the border must be moved to the middle by taking it out from the box. It can be done (assuming no friction) without any work done and no heat generated. In this transition, the knowledge on the position of the particle is lost, and therefore, the demon's memory is no longer correlated with its state. Therefore, it can be deleted in order to restore initial conditions for the new cycle.}
\end{enumerate}

Since the system operates in the cycle the Second Law of Thermodynamics, i.e.,
\begin{Theorem}\textbf{The Second Law of Thermodynamics (Kelvin) \cite{GeometryOfPhysicsFrankel}} \\
In the quasi-static cyclic process, a quantity of heat cannot be converted entirely into its mechanical equivalent of work.
\end{Theorem}
is valid. However, the system does not return any heat to the environment, and therefore the Law is broken. In order to force the system to obey the law, it must return to the environment the amount of heat that is no less than the work done, i.e., $Q \geq k_{B}Tln(2)$. As it was pointed out by Bennet \cite{Bennet, BennetDemon}, this is precisely the Landauer's bound and can be associated with erasing the demon's memory in the transition $5$ in Fig. \ref{Fig.SzilardVersion}.

This experiment can also be seen as the thermodynamical implementation of the computer memory, and we will use it in the next section to describe the details of the calculation of Landauer's bound for higher than binary systems. Although the specific thermodynamical system was selected, the final result will not depend on specific characteristics of this particular system, which suggests that the result is universal.

\section{One-hot encoding and the Landauer's bound for memory of $B$-ary system.}
The $B >0$ values can be uniquely coded in binary system using so-called one-hot encoding. This method is currently used in Machine Learning for encoding categorical values \cite{Raschka}.  Due to this coding, the alphabet is the set of digits/letters of $B$-system and the resulting information is unique code of each element of this alphabet in the $B$ bits. For $B$ values (from $0$ to $B-1$) such encoding will be as follows
\begin{equation}
 \begin{array}{ccc}
  b_{1}=0 & \Leftrightarrow & \underbrace{0\ldots 0}_{B-1}1 \\
  b_{2}=1 & \Leftrightarrow & \underbrace{0\ldots 0}_{B-2}10 \\
  {} & \ldots & {} \\
  b_{B}=B-1 & \Leftrightarrow & 1 \underbrace{0\ldots 0}_{B-1}. \\
 \end{array}
 \label{Eq.rangeOne-Hot}
\end{equation}
We can also add the value $b_{0}=\underbrace{0\ldots0}_{B}$, which does not belong to the encoding, but we add it since it will be used as an undetermined value. The coding is one of the worst, since there is abundance of digits, and since the numbers with two, three etc. $1$'s,  are not used. The efficiency of this coding (see \cite{InformationEntropy}, equation (4-2)) is
\begin{equation}
 E=\frac{log_{2}(B)}{log_{2}(2)}=log_{2}(B).
 \label{Eq.EfficiencyOfCoding}
\end{equation}
This quantity puts the bound on how many bits is necessary to encode the information.

\begin{figure}
\centering
 \includegraphics[width=0.5\textwidth]{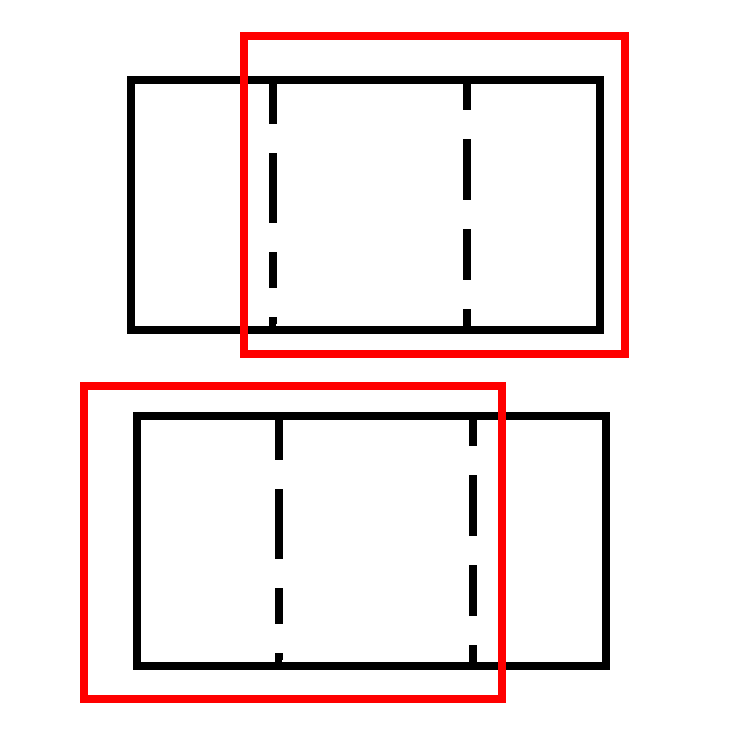}
 \caption{Minimal expansion for trit. The chambers which take part in EXTRACT part (closest neighbors) are marked in red.}
 \label{Fig.MinimalExpansion}
\end{figure}

Consider now the following experiment with a one particle of ideal gas in the box, similarly to the Szilard's idea. The cycle consists of the following steps:
\begin{enumerate}
 \item {START: Put $B-1$ borders equidistantly\footnote{If we would use non-equal volume splitting into $B$ parts, then averaging of emitted heat for each configuration must be used - this reflects tha fact that the Landauer's principle results from statistical considerations and averaging \cite{Piechocinska, LandauerExplainedFull} on the most fundamental level of statistical physics.} inside the parallelepiped box of volume $V$, that we get $B$ chambers of volume $V/B$.}
 \item {LOCALIZE: Localize the particle in the chamber encoded by a $1$ in sequence, e.g.,  $b_{i}$.}
 \item {EXTRACT: Start to expand isothermally the particle from the chamber of volume $V/B$ to the selected ONE nearest chamber finishing with the particle in the chamber of volume $2V/B$. This allows to extract the work of this minimal expansion
 \begin{equation}
 W_{min}=\int_{V/B}^{2V/B} k_{B}T\frac{dV}{V}=k_{B}T\ln(2).
 \label{Eq.MinimalDecompression}
 \end{equation}
 }
 \item {RESET: Reset the system by taking out all borders from the box that particle can again move freely. Then put the borders into the box equidistantly again and reset the memory of $b_{i}$ and put the particle again. It is irreversible free expansion with no work and no heat generation or consumption.}
\end{enumerate}
The situation for trit is presented in Fig. \ref{Fig.MinimalExpansion}.

In this approach, everything depends on which base the memory that stores the information on the particle localization is constructed.

For binary memory, we have a full one-hot encoding of the numbers $\{b_{1},\ldots, b_{B}\}$. In this case, in the EXTRACT step, we can trace the location of the particles up to the $V/B$ volume, in the sense that if for trit ($B=3$) the initial data was $b_{1}=0=001_{2}$, then after expansion it will be coding $a=011_{2}$, which is not allowed, since in the one-hot encoding range (\ref{Eq.rangeOne-Hot}). Then resetting the cycle is equivalent to the erasure in binary Szilard experiment described in the previous section, and therefore, generates the Landauer's bound for heat.

For memory based on the system with $B$ states ($B$-it), the EXTRACT part gives the state which cannot be described by the single number from the $B$-it system. Therefore RESTORE part reset the system and memory, giving the same Landauer's bound for erasing a single $B$-it for the base $B$ system. 

This is an alternative approach to the argument presented in \cite{TritLandauer}. It also explains the minimality of Landauer's bound. The above discussion shows that if the memory cannot store more coarse information about the state of the system, then the Landauer's bound for heat must be calculated concerning the minimal change of the system that cannot be stored in the memory. If we expand the initial chamber to the whole volume $V$ then the work is $W_{max}=k_{B}T\ln(B)$ and this is also a minimal bound for the heat in this experiment (resulting from the Second Law of thermodynamics), which is higher than the Landauer's bound. To our knowledge, this observation was not made before for $B>2$, even though it is a simple derivation of the case $B=2$. 

We can also consider more substantial changes in the system using more than two neighbor cells. This will be considered in the next section.

\section{More general decompression}
\begin{figure}
\centering
 \includegraphics[width=0.5\textwidth]{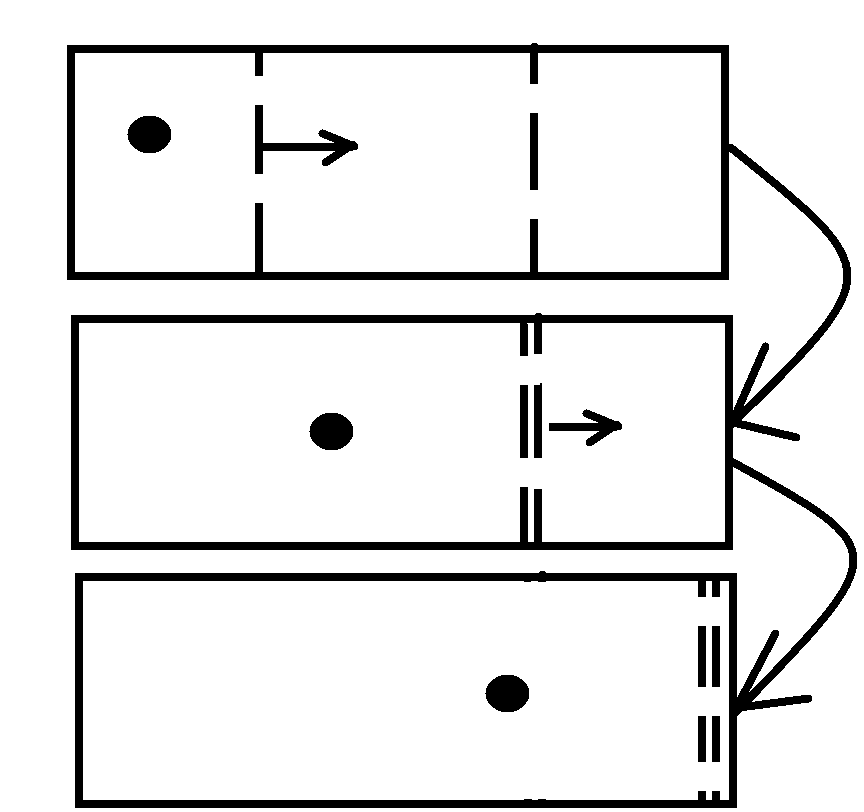}
 \caption{The case of 'Maximal' expansion both pistons are shifted to the boundary and the gas makes a work $W=k_{B}T\int_{V/3}^{V}\frac{dV}{V}=k_{B}T ln(3)$, which is bigger and needs the same or larger compensation by expelled heat in order to preserve the 2nd Law of Thermodynamics.}
 \label{Fig.MaximalExpansion}
\end{figure}

Consider now the same experiment as before with the expansion of the single ideal gas particle to the $2< N \leq B$ neighbor chambers. For trit, there is only one such expansion for $N>2$, and it is 'maximal'. It was presented in Fig. \ref{Fig.MaximalExpansion}.

In this case the particle expanding isothermally from $V/B$ initial volume chamber to the $NV/B$ final volume makes the work
\begin{equation}
 W_{N}=k_{B}T\int_{V/B}^{NV/B}\frac{dV}{V}=k_{B}T\ln(N),
\end{equation}
which is independent of $B$. This work also puts a minimal bound for the heat of erasure of the information by expanding to $N$ chambers and is required by the Second Law of Thermodynamics. Comparing this to minimal decompression (\ref{Eq.MinimalDecompression}) we get the increase of work we did due to additional (non-minimal/non-optimal for coding) decompression
\begin{equation}
 E_{N}=\frac{W_{N}}{W_{min}}=log_{2}(N),
 \label{Eq.RatioWork}
\end{equation}
which is also effectiveness of coding of the $N$ different values using binary coding. For example for $N=3$ (and $B=3$) we get the result of \cite{TritLandauer}. However in general, it can be used for $N=3$ and $B>3$. 

In this context, the correction factor $log_{2}(3)$ in \cite{TritLandauer} for efficiency of encoding is coincidence since $B=N=3$ for trit - the equation (\ref{Eq.EfficiencyOfCoding}) is then the same as (\ref{Eq.RatioWork}). In general, the number of individual chambers $B$ can be larger than the number of chambers which are merged in decompression. This remark will be even more visible when we discuss the implementation of the memory in the next section.

\section{Thermodynamical memory}
In this section, the thermodynamic realization of memory will be presented. It is provided here to avoid indirect derivation of Landauer's bound from the previous chapter and possible 'circular argument' accusations. The optimal implementation that reaches Landauer's bound and then non-optimal implementations will be presented. We will also provide a connection of these implementations with the Galois connection associated with memory systems in \cite{KyciaLandauer, KyciaLandauer2}.

\subsection{One-hot optimal implementation}
\begin{figure}
\centering
 \includegraphics[width=0.5\textwidth]{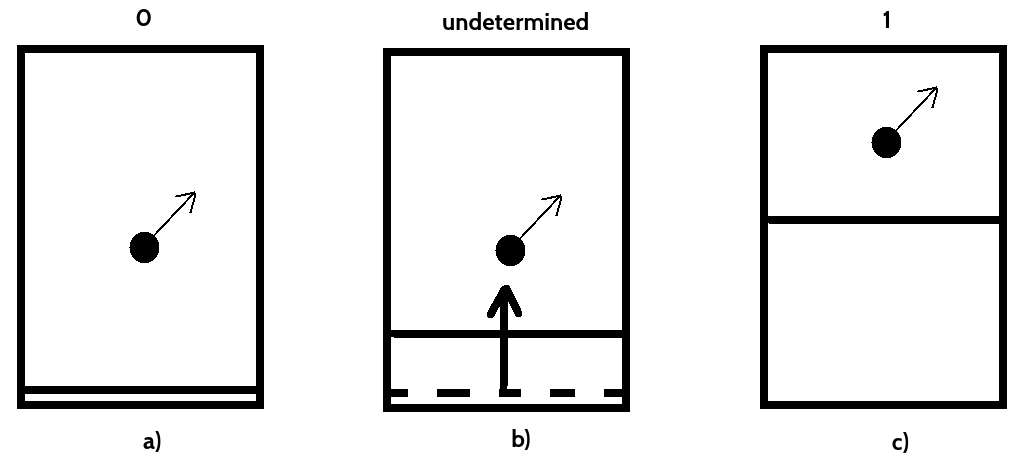}
 \caption{An implementation of a single bit of a memory. It consists of a box with movable border and a single particle of a gas in thermal bath of temperature $T$. a) state encodes $0$, b) represents isothermic compression from the volume $V$ to $V/2$ and represents undetermined (internal) state of memory, c) represents the state $1$.}
 \label{Fig.OneHotBit}
\end{figure}

This implementation is based on one-hot encoding. First, the single bit is constructed by reversing the construction from the Szilard's description. The situation is presented in Fig. \ref{Fig.OneHotBit}. It is a box with movable border and a single particle of an ideal gas in thermal contact with themostat (environment) of temperature $T$. There is an additional source of energy that powers the memory which is not visible in the figure.
The representation of a bit is as follows:
\begin{itemize}
 \item {a) - zero bit $0$;}
 \item {c) - one bit $1$;}
\end{itemize}
and the transitions:
\begin{itemize}
 \item {b) - isothermal (reversible) compression from the volume $V$ to $V/2$. The heat expelled to the environment is $Q=-k_{B}T\int_{V}^{V/2}\frac{dV}{V}=k_{B}T\ln(2)$.}
 \item {isothermal (reversible) decompression(opposite to  b)) from c) to a), which extracts the heat $Q=k_{B}T\ln(2)$ from the environment and converts it to work.}
 \item {transition from c) to a) that results from adiabatic free decompression that is realized by taking out from the box the border from the middle and put it to the box along one of the border of the box. It is free (irreversible) decompression with no work and heat generated or consumed.}
\end{itemize}

As was pointed out in \cite{KyciaLandauer}, the implementation of the memory can be associated with the Galois connection. It is the relation between two sets\footnote{Generally, the Galois connection is a relation between two ordered categories, which are not necessary sets, however in this presentation, we restrict ourselves to less general Galois connection between ordered sets.} that have some ordering relations of elements - they are called pre-ordered sets, or posets for short. Then the Galois connection consists of two maps between sets in opposite directions, which preserve this ordering.  It is a prototype of more general connections between two categories called adjointness \cite{CategoryGentleIntroduction, SpivakSketches}. For physical applications it is probably too weak notion (it is not isomorphism) to be useful, however, it has an interesting interpretation which shed some light on its occurrence in this context: It represents the relation between theories and their implementations on the model\footnote{One poset is a set of theories ordered by finer assumptions, and the second poset is set of models for these theories also ordered by the finer details. See \cite{CategoryGentleIntroduction, SpivakSketches} for details.}, or differently, between abstraction map (from model to theory) and realization map (from theory to model). This scenario is evidently present in considerations of implementing Boolean algebra on a physical memory device. There are two levels - logic and physical system on which the logic is implemented by labeling specific configurations as it will be presented hereafter.

\begin{figure}
\centering
 \includegraphics[width=0.5\textwidth]{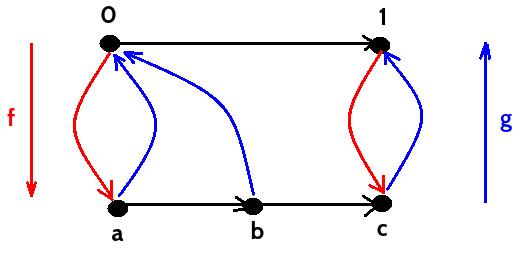}
 \caption{Mapping between logical states and the states of physical realization of memory. The arrows between posets states are given by its ordering, i.e., since $0\leq 1$, so there is $0\rightarrow 1$, which is usual convention for treating a poset as category on its own \cite{CategoryGentleIntroduction, SpivakSketches}.}
 \label{Fig.GaloisBit}
\end{figure}

Introducing the ordering at the level of bits $0 \leq1$ we get a poset $A$. At the level of physical implementation of the bit, we have ordering $a \leq b \leq c$ we get a poset $B$. Then the implementation of binary logic in physical memory can be presented by the mapping (functor) $f: A \rightarrow B$, and the mapping (functor) $g:B\rightarrow A$ presented in Fig. \ref{Fig.GaloisBit}.
The map fulfils the following Galois connection condition \cite{KyciaLandauer, CategoryGentleIntroduction, SpivakSketches}
\begin{equation}
 f(p) \leq q \quad \Leftrightarrow \quad p \leq g(q),
\end{equation}
for $p \in A$ and $q \in B$. That is $f$ is left adjoint to $g$, i.e., $f \dashv g$ and not the otherwise as it can be checked by inspecting all possible pairs $p,q$.

In addition, the transitions can be
\begin{itemize}
 \item {$a\rightarrow b \rightarrow c$ - reversible isothermal compression operation associated with logically reversible (bijective, see \cite{LandauerExplainedFull}) NOT: $0\rightarrow 1$. Heat emitted to the environment $Q = k_{B}T\ln (2)$.}
 \item {$c\rightarrow b \rightarrow a$ - reversible isothermal decompression operation associated with logical reversible NOT: $1\rightarrow 0$. Heat absorbed $Q = k_{B}T\ln (2)$.}
 \item {$c \rightarrow a$ - irreversible free decompression operation associated with logical irreversible deletion $x \rightarrow 0$.}
\end{itemize}
As it can be noted, if two times NOT reversible operation is applied, then there is no net heat emitted or absorbed. However, if reversible NOT and then deletion will be performed then the net heat, equally as  Landauer's bound, will be emitted as required by Landauer's principle or the Second Law of Thermodynamics. This observation was generalized in the following Table \ref{Tab.LandauerConnection} (see \cite{LandauerExplained, KyciaLandauer}),
\begin{table}
\centering
\begin{tabular}[!h]{|c|c|c|}
 \hline
 Possibilities &  \begin{tabular}{@{}c@{}}$B$ \\ reversible\end{tabular}  & \begin{tabular}{@{}c@{}}$B$ \\ irreversible\end{tabular} \\ \hline
 $A$ reversible & YES  & YES \\ \hline
 $A$ irreversible & NO & YES \\ \hline
\end{tabular}
\caption{Systems $A$ is implemented on $B$.} 
\label{Tab.LandauerConnection}
\end{table}
which explains which types of operations can be realized between both Galois (Ladauer's) connected systems.

This single-bit memory cell can be composed to implement a one-hot encoding of the state of the extended Szilard minimal expansion experiment presented in the previous section. Since the deletion of the memory in this experiment is connected with one bit, therefore, the Landauer's bound is achieved.

The Galois connection is represented in this case introducing partial order in one-hot encoding set $b_{0} \leq b_{i}$ for $i=1\ldots N$ that makes the poset $A$. At the level of memory implementation, the ordering is induced by the ordering for a single bit ordering from Fig. \ref{Fig.GaloisBit}, since only one bit in memory is changing.

\begin{figure}
\centering
 \includegraphics[width=0.5\textwidth]{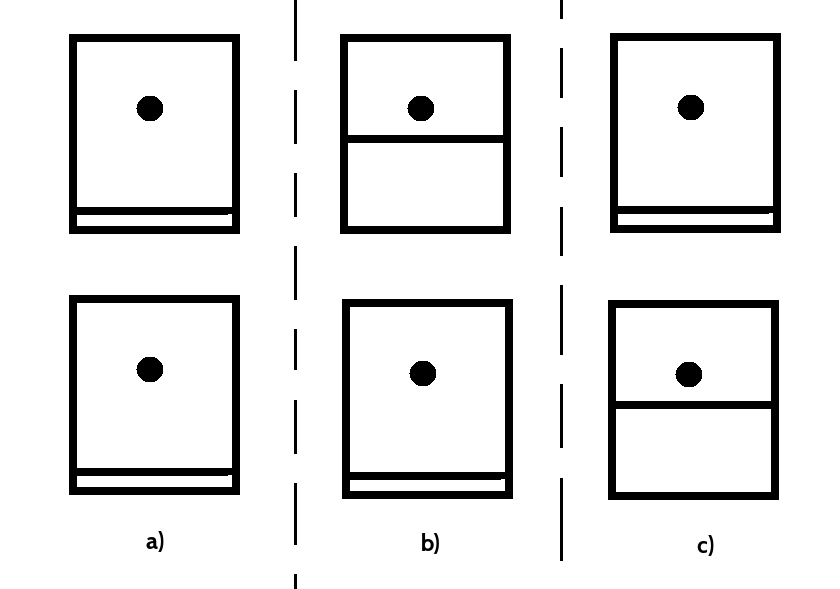}
 \caption{One-hot encoding of a single bit thermodynamical model: a) represents undefined state; b) represents, say, $1$; c) represents $0$ bit. Reversible operation $0 \longleftrightarrow 1$ can be made by making the pair of isothermal decompression-compression sequence on complementary up-down chambers that pass trough a) state. An irreversible operation engages free decompression to a) state and then a transition to the required configuration ($0$ or $1$) by isothermal compression, which generates Landauer's heat bound. The ordering is $b \leq a \leq c$ (in more details $a \leq b$, $a \leq c$ and $b \leq c$) which corresponds to the ordering $0 \leq 1$ on logical level.}
 \label{Fig.TrueBit}
\end{figure}

A final remark on the real implementation of the optimal memory. In Szilard's version of the experiment, only a single bit (a single particle) must be traced, and therefore $0$ bit can describe the undetermined state or no particle state. However, when both $0$ and $1$ bits have some meaning in the experiment, then the single bit must be implemented, as in Fig. \ref{Fig.TrueBit}.

\subsection{Non-optimal vs optimal implementation}
We propose the implementation of thermodynamical memory, which during (irreversible) erasure, generates more heat than Landauer's bound. It can be made trivially using the one-hot implementation from the above subsection by using irreversible compression instead of isothermal (reversible) one. This is off-diagonal 'YES' case in Table \ref{Tab.LandauerConnection}.

Another possibility and this will be an elaboration of the idea noted in \cite{TritLandauer}, is by a reversal of non-minimal expansion Szilard's version of the experiment. In this approach, the memory has some abundance of internal states not used in the representation of bits. This makes that the Landauer's bound is not reached.

\begin{figure}
\centering
 \includegraphics[width=0.6\textwidth]{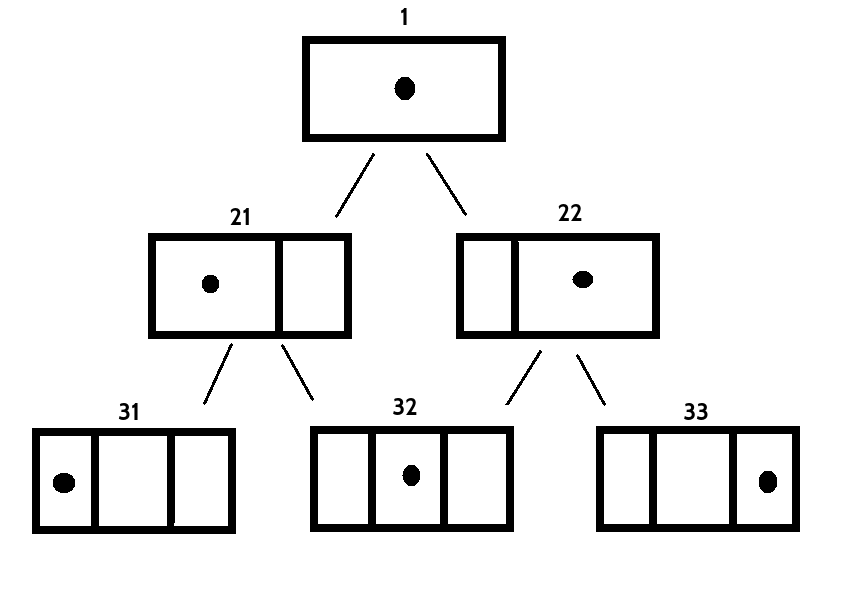}
 \caption{Trit implementation involves mapping $31 \rightarrow 1$, $32 \rightarrow 2$ and $33 \rightarrow 3$. The internal states $21,22,1$ do not represent any trit and are used to transitions between elements. The ordering is made by the transitions along the graph, e.g., $31 \leq 21 \leq 32$. Trit ordering is as usual $1 \leq 2 \leq 3$.}
 \label{Fig.TritNonOptimal}
\end{figure}

\begin{figure}
\centering
 \includegraphics[width=0.6\textwidth]{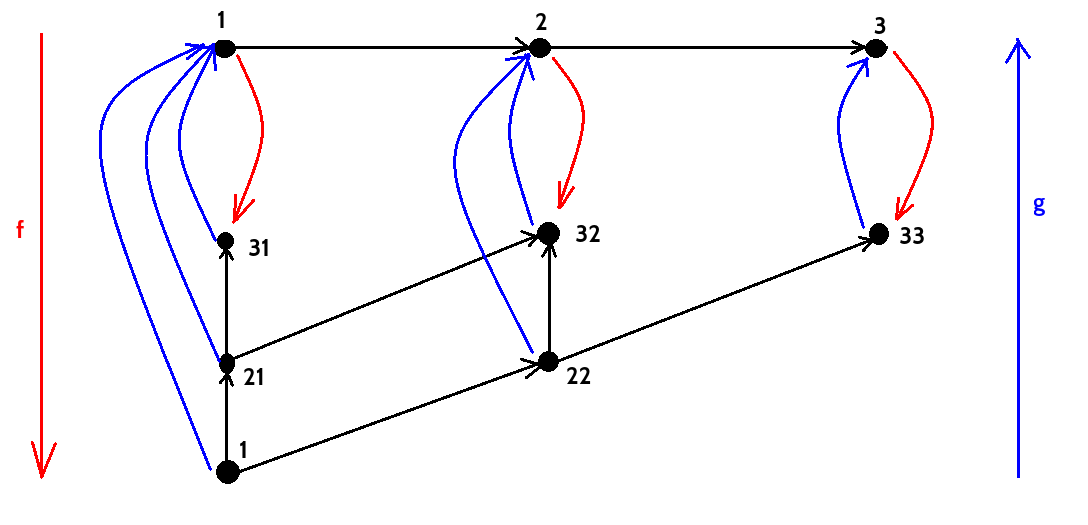}
 \caption{Galois connection $f \dashv g$ between poset of trit logic and physical implementation of memory using Szilard's approach. The poset at the level of physical implementation can be constructed using an ordering ('reversed' inclusion, e.g. $111 \rightarrow 110$) in the Boolean algebra of three bits via mapping $1 \rightarrow 111$, $21 \rightarrow 110$, $22 \rightarrow 011$, $31 \rightarrow 100$, $32 \rightarrow 010$, $33 \rightarrow 001$. Here $1$s can be associated with the knowledge of which chambers can the particle penetrate, and $0$s describe which chambers are not available to the particle.}
 \label{Fig.GaloisTritNonOptimal}
\end{figure}

An example for the trit is presented in Fig. \ref{Fig.TritNonOptimal}.
For completeness, Galois connection is presented in Fig. \ref{Fig.GaloisTritNonOptimal}.
This idea can be extended to arbitrary base $B>1$.

Consider an irreversible free decompression $31 \rightarrow 21$ and then reversible operation $21 \rightarrow 32$, which corresponds to logical irreversible transition $1\rightarrow 2$. Irreversible operation $31 \rightarrow 21$ generates no heat, however isothermal compression $21 \rightarrow 32$ from volume $2V/3$ to $V/3$ generates the heat $Q = k_{B}T \ln(2)$, so the Landauer's heat is attained and this implementation is optimal too. It is even less complex than one-hot encoding.

Consider however the same transition that engages irreversible free decompression $31 \rightarrow 1$ and then isothermal compression $1 \rightarrow (21 \vee 22) \rightarrow 32$ from volume $V$ to $V/3$, which generates the heat $k_{B}T \ln(3)$. This path engages the states hidden on a deeper level that the layer neighbor to the one giving the representation of digits for trit, and gives the heat greater than the Landauer's bound. 

This example shows that the optimal implementation of the memory can be done by reversing Szilard's version of Maxwell's demon using minimal decompression. Besides, the more levels of the tree (an example of which is Fig. \ref{Fig.TritNonOptimal}) the transition engages, the larger the bound for the heat expelled. The correction factor is exactly the ratio of the depth of the levels of the tree used to implementation of memory to the optimal implementation (minimal decompression), for trit it is $\frac{\ln(2)}{\ln(3)}$, however for the larger base, the denominator can be larger. For trit (see \cite{TritLandauer}), there is only minimal ($\log_{2}(2)$) and maximal ($\log_{2}(3)$) levels to use in transitions, and therefore the ratio $\frac{\ln(2)}{\ln(3)}$ is unique. However, for larger base the correction factor can range from $\frac{\ln(2)}{\ln(3)}$ to $\frac{\ln(2)}{\ln(B)}$, which is a caveat or pitfall to derivation of Landauer's bound for general-base memory.

The Landauer's bound is always true, however, given memory can have higher bound for heat emission for irreversible operations and never reach the Landauer's bound.

The above example also shows that the irreversible decompression makes it impossible to restore the work previously stored in the heat of the environment, and this is the main idea behind the Landauer's bound. The reversible processes use this heat to change the state by reversible isothermal compression-decompression transition between top states. From this viewpoint, the information is encoded in the ability to store and restore energy from the system which is treated as an operation on the information. Therefore the energy transitions in memory are a fundamental level (physical realization) on which the information level is constructed.

\section{Summary}

We have shown, using Szilard's version of Maxwell's demon experiment, the validity and difficulties in interpretation of Landauer's bound. The essential ingredient in the implementation is an engagement of the inner states of the memory. The Landauer's bound is always true, however, some memory systems have a higher value of the bound for heat emission during the irreversible operation due to a more complex structure of internal states of the memory. It is hoped that this will be an additional argument to the statement that all our models are accurate up to some scale, and the second law of thermodynamic governs what happens with these inaccuracies.

The one-hot encoding implementation of memory, as well as the implementations using reversed Szilard's version of Maxwell's demon experiments, show problems and bounds resulting from the Second Law of Thermodynamics for these realizations. It also shows that the Second Law of Thermodynamics is more fundamental than Landauer's principle on the level of classical (equilibrium) Thermodynamics - by specific choice of implementation of memory one can never reach Landauer's bound even without friction in the system.

We also presented relations to abstract Galois connection that relates implementation of memory with the base system that is stored in it. Appearance of this theoretical construction is expected in every situation when we have two systems when one can be considered as an abstract theory (Boolean algebra or multivalued logic) and the second is its realization/implementation(physical implementation of memory).

\section*{Acknowledgments}
RK was supported by the GACR grant GA19-06357S and Masaryk University grant MUNI/A/0885/2019. We would like to acknowledge COST CA18223 action for support.

\appendix


\end{document}